\documentclass[12pt,preprint]{aastex}
\setkeys{Gin}{width=0.85\textwidth}
\usepackage{graphicx}

\newcommand{\hmpc}{\ifmmode{h^{-1}\,\hbox{Mpc}}\else{$h^{-1}$\thinspace Mpc}\fi}

\newcommand{\kms}{\ifmmode{\,\hbox{km\,s}^{-1}}\else {\rm\,km\,s$^{-1}$}\fi}
\newcommand{\msun}{{\rm\,M_\odot}}

\begin{document}
\title{Modeling GD-1 Gaps in a Milky-Way Potential} 
\shorttitle{GD-1 Gaps}
\shortauthors{Carlberg}
\author{R. G. Carlberg}
\affil{Department of Astronomy \& Astrophysics, University of Toronto, Toronto, ON M5S 3H4, Canada} 
\email{carlberg@astro.utoronto.ca }

\begin{abstract}

The GD-1 star stream is currently the best available for identifying density fluctuations, ``gaps", along
its length as a test of  the LCDM prediction of large
numbers of  dark matter sub-halos orbiting in the halo. 
Density variations of some form are present, since the variance of the density along the stream is three times that
expected from the empirically estimated variation in the filtered mean star counts.
The density variations are characterized with filters  that approximate the shape of
sub-halo gravitationally induced stream gaps. The filters locate gaps and measures their amplitude,
leading to a measurement of the distribution of gap widths. To gain understanding
of the factors influencing the gap width  distribution, a suite of collisionless n-body simulations  
for a GD-1 like orbit in a Milky Way-like potential
provides a dynamically realistic statistical prediction of the gap distribution. 
The simulations show that every location in the stream has been disturbed to some degree by a sub-halo.
The small gaps found via the  filtering are largely noise.
Larger gaps,  those  longer than 1 kpc, or 10\degr\ for GD-1, are the source of the excess variance.
The suite of stream simulations shows that sub-halos at the predicted inner halo abundance or possibly somewhat higher
can produce the required large sale density variations.
\end{abstract}

\section{INTRODUCTION}
\nobreak

The GD-1 star stream \citep{GD:06} 
is, at 70 pc width,  one of the thinnest known. The implied low velocity dispersion,
about 1 \kms, means that GD-1 will respond coherently to the passage of 
the many low mass dark matter sub-halos predicted to present in the galactic dark matter
 halo in an LCDM cosmology \citep{Aquarius,VL1,Stadel:09}.
A dark matter sub-halo passing very near of through the streams causes a folding \citep{Carlberg:09} which 
observationally appears as density peaks on either side of a
low density gap \citep{YJH:11,Carlberg:12,Erkal:15}.  
Assuming an  LCDM cosmology allows concrete, statistical, predictions
of the properties of the  induced irregularities in star streams which can 
be compared to observational data.  There are alternatives to LCDM cosmology 
which work well on galactic scales and do not predict large numbers of unseen dark matter sub-halos \citep{Kroupa:15}. 

GD-1 is visible over about 80\degr\ of high latitude sky where extinction is low. 
Since the visible section of the  stream is a relatively nearby 8 kpc,  SDSS \citep{SDSS} imaging data 
reach the turn-off stars where the luminosity function rises steeply. 
\citet{GD:06} estimated 1800$\pm$200 stars were visible in the  63\degr\ of stream they examined.
Radial velocities  \citep{Yanny:09} and mean proper motions \citep{Munn:04,Munn:08},
provide some stream averaged phase space information \citep{Willett:09,KRH:10} along the stream.
The derived orbits have  a perigalacticon of about 14 kpc and apogalacticons of  26-29 kpc.
No progenitor for the GD-1 stream has yet been identified close to the orbit of the stream, although
the very low velocity dispersion of the stream strongly suggests that the progenitor is (or was) a low
mass globular cluster.

\citet{CG:13} characterized the gaps within GD-1 using 
a set of gap shaped filters that are convolved with 
the density along the stream.  The outcome is the identification of the locations and sizes of gaps.
The result for GD-1 was a distribution of gap sizes that behaved about as the predicted power law for gaps 
larger than 2\degr\ (using a width defined with the negative part of the gap) and flattened off for smaller gaps. 
The paper noted that the internal velocity dispersion and shear in the
stream suppressed small gaps  below the numbers predicted from
a  cold stream model, which was quantified to some
degree in \citet{Carlberg:15a}.

The purpose
of this paper is to compare the distribution of gap sizes in GD-1 to the same
measurements of dynamically realistic models of the GD-1 stream.
The stream progenitor will be assumed to be a low density globular 
cluster  on a GD-1 like orbit in a potential that approximates the Milky Way and contains the predicted
 level of dark matter sub-halos. A set of simulations is undertaken since the 
gaps in any one stream
depend significantly on the random variations in the sub-halo population as realized from the same
parent population of sub-halos.
The resulting streams are projected onto the sky and analyzed with the gap filtering technique,
which is applied in exactly the same way to 
the  observational data. The goal is  to quantitatively understand the 
numbers of GD-1 detected gaps as a function of their size as a test for consistency 
with the LCDM sub-halo predictions.

\section{Gaps in The GD-1 Stream}

\begin{figure}
\begin{center}
\includegraphics[angle=-90,scale=0.8]{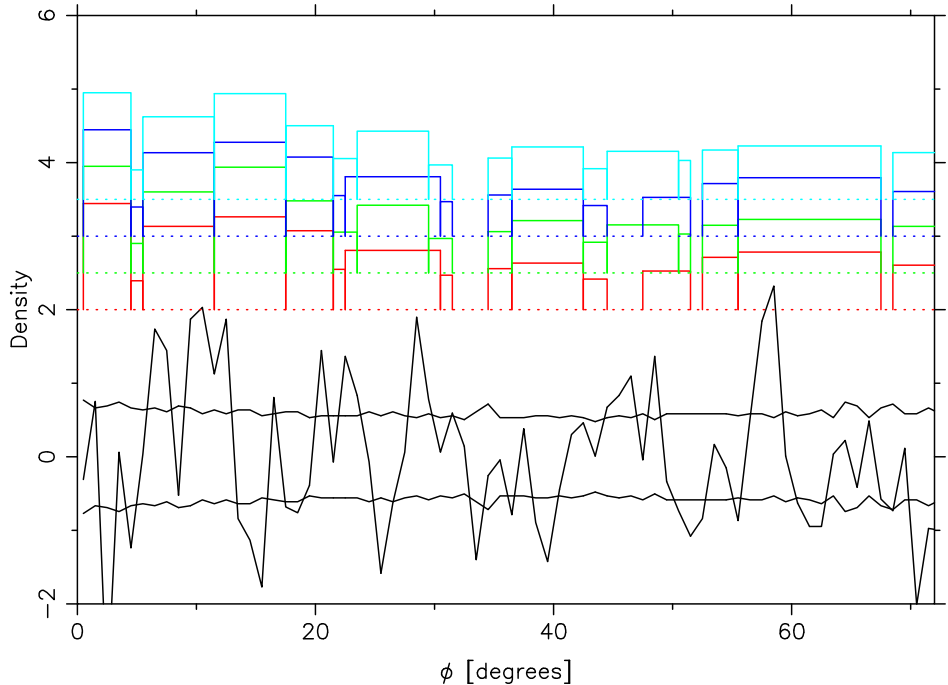}

\end{center}
\caption{The density along the GD-1 stream (jagged line) from the measurements of
\citet{CG:13} re-binned to 2\degr\ and normalized to a mean of one which is then subtracted. 
The accompanying pair of lines show the locally 
estimated $\pm 1\sigma$ standard deviation of the
density. The histograms are the gaps recovered from this stream data in the gap-finding
procedure discussed in the text.  
The gaps here include the compensating high densities on either side so the
gaps here are defined to be three times the width used in \citet{CG:13}.
Starting from the bottom, the
histograms are the results for the $w_1$ (red) and $w_2$ (green) symmetric filters, and then the same filters
augmented with their
asymmetric versions, blue and turquoise, respectively.}
\label{fig_gd1}
\end{figure}

The density of stars along the GD-1 stream was measured \citep{CG:13} using the SDSS DR-8 photometric data \citep{DR8}.   
Their filtered density of stars along  the stream, rebinned to 2\degr\ intervals, is shown in Figure~\ref{fig_gd1}.
Foreground and background stars along the line of sight to GD-1 are greatly, but far from completely, suppressed with a
 matched filter designed to pick out old metal poor stars in
stellar color-magnitude data \citep{Rockosi:02,Grillmair:09,Grillmair:11}. 
The mean over-density of  the filtered stream is 14\%  above  the 
 filtered mean background density.

The errors in the stream density measurement  is estimated from the variance in the field star density
near to the stream., augmented with the slight extra noise that the mean stream 
itself will contribute. The resulting signal-to-noise ratio per 2\degr\ bin is 2.3. 
However,  the variance in the stream density is more than three times
what is expected for a unstructured, but noisy, stream and is strong statistical evidence that the stream 
contains some form of excess density variations.

\begin{figure}
\begin{center}
\includegraphics[angle=-90,scale=0.8]{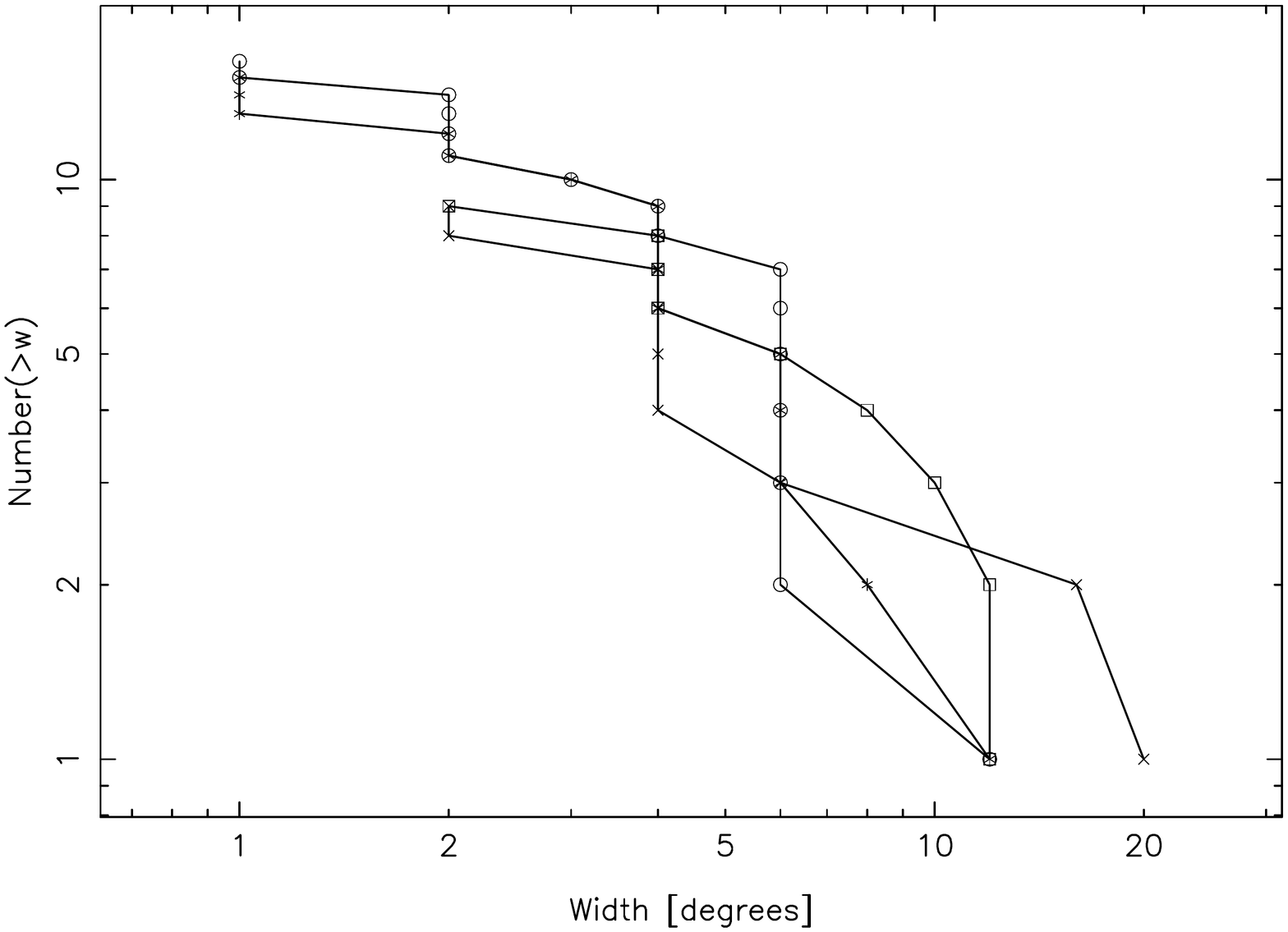}
\end{center}
\caption{The cumulative number of gaps as a function
of their width and filter shape used. The symbols are, 
for 1 degree bins:  asterisk, circle, and for 2 degree bins: cross, square, where the first and second member
of each pair is filter 1 and 2, respectively.}
\label{fig_ngd1}
\end{figure}

To characterize the excess variance of the stream density, 
we use a gap-filter designed to approximate the shapes of stream gaps \citep{Carlberg:13}.
The filters have a negative density in the center with two positive density peaks on either side. 
 The filters have zero mean over the range $x=[-3,3]$ and are normalized to unit variance.
Our preferred filter is
$ w_1(x)=(x^6 -1)\exp{(-1.2321 x^2)}$ 
  but as an alternate we use the spikier 
$ w_2(x)=(x^8 -1)\exp{(-0.559 x^4)}$, although a continuum of filter shapes should be examined at some future step.
As a first step towards recognizing that gaps are generally not symmetric \citep{Sanders:15}, 
we introduce some asymmetry with
a simple linear function, $a(x)=1+\alpha x$, to create a  right enhanced
asymmetric filter version 
of the $w_i$ filter,
$w_{ir}(x)=w_i(x) a(x), x>0$ and $w_{il}(x)/a(x), x<0$, with its left enhanced partner being the reversed version.
We use $\alpha=0.25$, however a whole continuum of $\alpha$ values could be examined. 
To prepare the density field for filtering, it is normalized to a mean of unity, which is then subtracted. 
We then pass the densities through a Gaussian high pass filter having a filter scale of half of  the fundamental wavelength
of 73\degr. 

Gap filtering transforms the one dimensional stream densities as a function of angular
distance along the stream, $d(\phi)$, into a 
two dimensional space of stream position and filter width, $D(\phi,g)$.  
The filter argument, $x$ is  $x=\phi/g$ where $g$ begins at the bin size.
The value of $g$ goes to 50\degr\ (which is larger than the maximum allowed gap in GD-1) in 100 logarithmic steps per decade, 
which creates 393 filters in total. Finer steps in filter size make no difference to the outcome and
even ten times coarser steps lead to relatively small differences in outcome.
Peaks in this two dimensional space are identified as points higher than the 
neighboring points, here simply the neighboring two points in $x$ and two in the values of $g$ (where $g$ may include asymmetric filters as well).
At any given $\phi$ value there are likely to be peaks at a number of differing $g$ values.
The height of the peak depends on how well matched the shape of the filter is to the density profile at that location
in the stream. A perfect match with no noise would give a peak height of one. 
Each filter is normalized to allow the
comparison of filters of different sizes. 
To place density bins  into a unique gap (or no gap at all) we 
start at the gaps found for the smallest $g$ value which identifies a series of peaks along
the stream. 
The points within width $\rm \pm 3g$ of each peak that is above some minimum threshold height into
a gap.   We then examine the next $g$ value upward in size. Any peaks of a longer
filter that are higher than for a lower filter takes over the bins within a shorter filter gap. This procedure
is continued to the largest $g$ value and leads to bins being placed into gaps of varying widths and heights.
A very low threshold for allowing a gap means that virtually the entire 
density field is in gaps with a very high threshold requiring an extremely good match to the prescribed shape of the gap.
We will turn to stream simulations for guidance on how to interpret
the gaps. 

Figure~\ref{fig_gd1} displays the results of the gap filtering applied to the GD-1 densities in 2\degr\ bins.
The gaps are defined as initially extending over  [-3g, 3g] not allowing any of the included region 
to have a gap unless it is a deeper gap.  Longer gaps can take over some or all of a shorter gap.
In detail these procedures are similar but not identical to those used in \citet{CG:13}.
The gaps  in \citet{CG:13} were defined to initially extend from [-g, g], roughly the negative density
region of the gap and  $1/3$ of the widths adopted here.
The gaps resulting from the revised analysis here are
 displayed with a vertical offset of 0.5 units at the top of the density plot, Figure~\ref{fig_gd1}.
The $w_1$ (red) and $w_2$ (green) filters largely lead to similar results. 
The same filters augmented with their left and right asymmetric versions, 
blue and turquoise, respectively, again give very similar results, 
with one notable large gap being preferred over several smaller gaps for the asymmetry allowed $w_2$ 
(turquoise, the top set of gaps).

Figure~\ref{fig_ngd1}  shows the cumulative number of gaps greater in length than some value $\rm w$, $N(>{\rm w})$ 
for the two symmetric filters, here run on the density field binned into both 1\degr\ and 2\degr\ bins. 
Overall the two filters and the two different data bin widths recover a similar underlying distribution of widths. However, 
the 1\degr\ binning has a large number of small gaps recovered. The excess of small gaps in
the 1\degr\ binned data over those in the
2\degr\ binned data is nearly a factor of 3 and it is reasonable to suspect the increased numbers of gaps at small scales
is largely due to noise.

\section{GD-1 Stream Simulations}

\citet{CG:13} compared similar, but not identical, gap measurements of GD-1 to predictions from approximate fits to 
a set of simulations of streams in a external potential with orbiting sub-halos
derived  from the  \citet{Aquarius}  LCDM numerical model halo. 
 A limitation of
that work is that
the theoretical analysis
examined only streams on circular orbits, where gaps are much simpler and persist long than a stream 
on an elliptical orbit. The sub-halo interactions with the stream
were done individually with the cumulative effect simply being the mass weighted sum over all sub-halos,
with no allowance for overlap \citep{NC:14}. 

Simulations  are essential to give
 the appropriate range of ages along the stream, 
the gap-blurring of velocity dispersion and velocity shear across the stream, 
and the internal structure of the stream which varies with orbital phase.  
The dynamical properties of the leading and trailing streams drawn out of a globular star cluster progenitor
through tidal heating 
can be accurately calculated 
within a collisionless n-body code \citep{Carlberg:15a,Sanders:15}. Allowing
two-body relaxation between stars within the cluster \citep{Kupper:10,Kupper:12} adds more precision 
for the stream properties of those clusters where energy equipartition processes are important,
but requires a more detailed description of the cluster and its contents and adds computational costs.

The Milky Way potential MWPotential2014 \citep{Bovy:15}, here referred to as MW2014, is a somewhat simplified 
(it has no bar, for instance), easy-to-use, potential 
 designed to provide a good description of the mean axisymmetric Milky Way potential
 from well inside the solar circle to 60 kpc, 
which more than covers the radial range that the orbit of GD-1 traverses. 
The assumption of an axisymmetric potential eliminates most chaotic orbit possibilities
\citep{PW:15}. 
The MW2014 dark halo component is described with an NFW 
model function \citep{NFW}. The scale radius of 16 kpc, very similar to the
value of $r_{200}/c_{NFW}=15.3$ kpc of the Aquarius 
numerical model \citep{Aquarius}.  
MW2014 has  an enclosed dark matter mass out to 245 kpc of $8.1\times 10^{11}\msun$,
44\% of the Aquarius mass to that radius, which proportionally reduces the total mass of sub-halos.

The model contains a set
of  sub-halos which interact with the stream to create gaps in a realistic manner.
Over a galactic lifetime the gaps can overlap, which is naturally taken
into account.  
The sub-halos are much less concentrated than the dark matter, 
with a radial distribution 
well described with an Einasto function \citep{Aquarius} where we use their overall fractional mass normalization, although
the total dark halo here is less massive than in their simulations.
The local mass fraction in sub-halos inside 16 kpc is 0.0014 with a fraction of 0.0038 inside 32 kpc, that is, 
sub-halos are
are very small fraction of the mass in the inner halo.
Out to 245 kpc, the $r_{200}$ radius of Aquarius, the mass fraction in sub-halos rises to 0.058.

The sub-halo distribution is assumed to be spherical, as is the overall dark matter halo. The 
Milky Way sub-halos surrounding dwarf galaxies are known to be in a fairly flattened distribution 
\citep{LB:76,Pawlowski:12,PK:14} 
with the distribution of the unseen sub-halos
of course not known. If they have the same distribution as the visible ones then the rate of
 interactions will depend on the orbit of GD-1
relative to the plane of satellites. \citet{Willett:09} find that the GD-1 orbit is inclined to the galactic plane about 35\degr, so 
the pole of the orbital plane is 55\degr\ from the galactic plane, currently in a direction away from the galactic center, but
the orbit precesses slowly around the galactic pole. 
\citet{PK:14} find a satellite plane pole at $\ell\simeq 159\degr$, $b\simeq -5$ and that the average 
distance from the pole to all satellite orbit poles is 48\degr. Therefore 
this crudely estimated GD-1 orbital pole lies within that angular distance, but near the outer edge of the cluster of points shown 
in their Figure~3. Modeling the plane of satellites as flattened sphere
finds a $c/a\simeq 0.25$ for a sample including lower
mass dwarf galaxies \citep{Pawlowski:16}. Therefore the 
interactions would enhanced up to a factor of four if the stream lay exactly in the plane and less with increasing misalignment. 
This is a potentially interesting effect which merits further investigation.

\begin{figure}
\begin{center}
\includegraphics[angle=-90,scale=1.1]{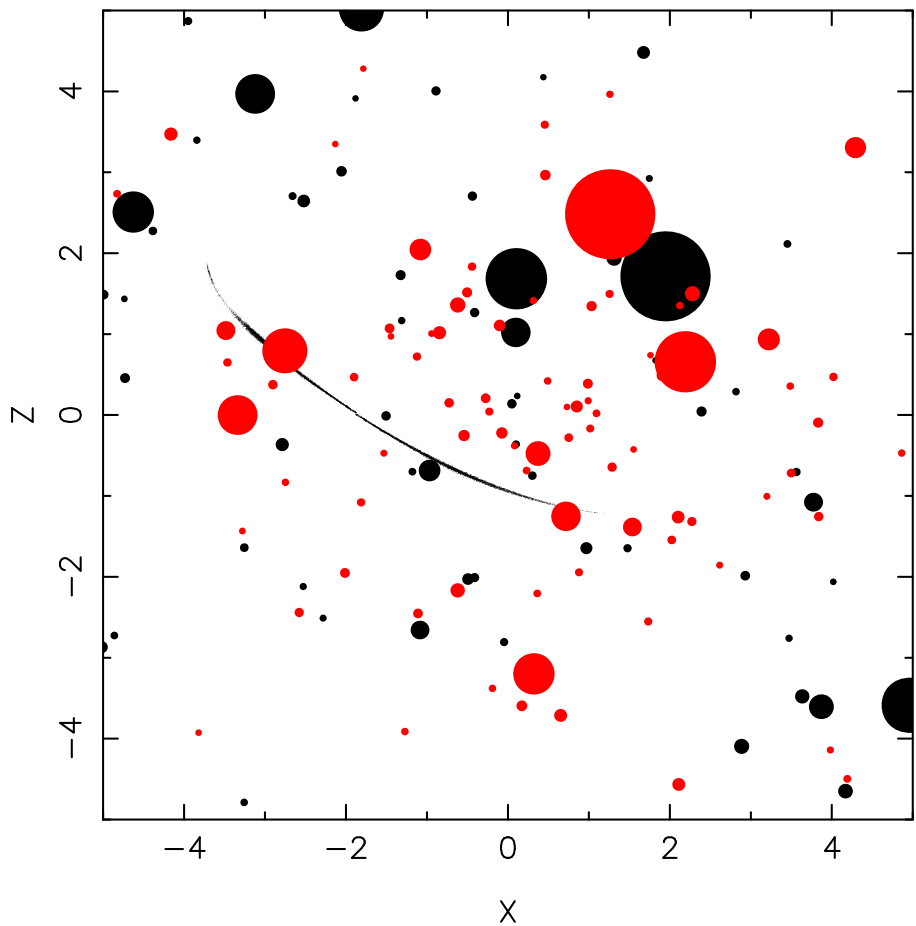}
\end{center}
\caption{The edge-on projection of the distribution of sub-halos at the beginning (black) and end of 
a simulation,  10.44 Gyr later (red), with 
the stream shown at the end of the simulation. 
The sub-halos are  given a size 5 times their scale radius.
A density weighted version of the stream would
appear thinner. One unit of distance is 8 kpc. The solar location is
in the galactic plane, $z=0$, at $x=1$.
}
\label{fig_subhalos}
\end{figure}

\begin{figure}
\begin{center}
\includegraphics[angle=-90,scale=0.9]{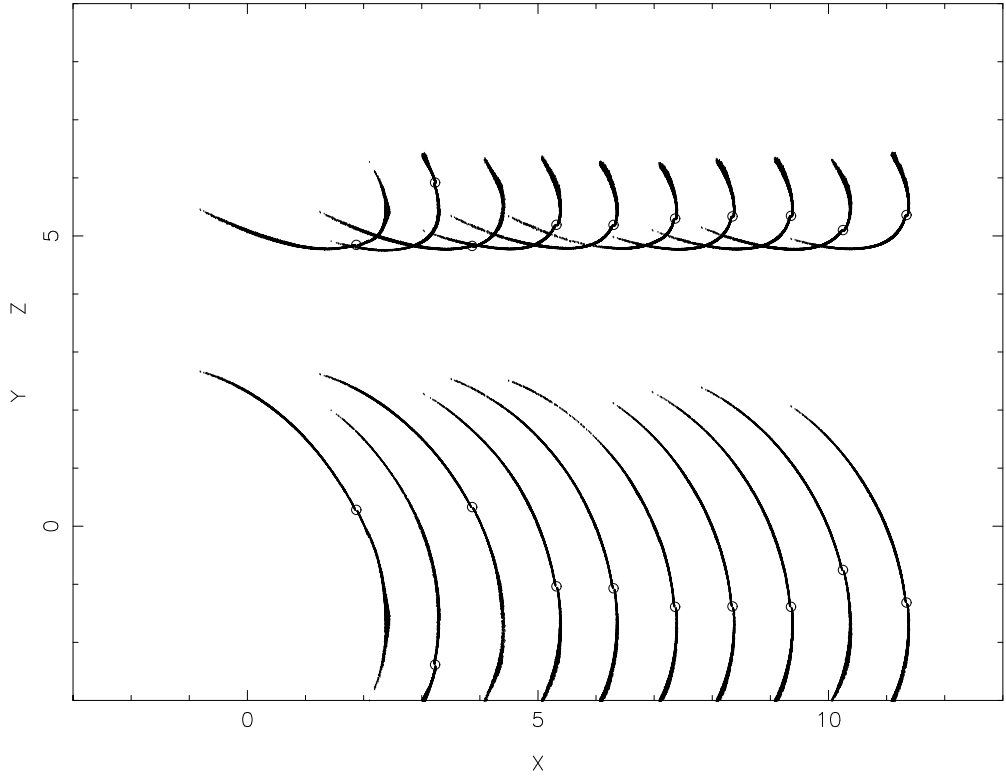}
\end{center}
\caption{$xy$ (lower row) and $xz$ (upper row) plots for ten realizations
of the same initial conditions.  The streams are all viewed at the same time, 293.6,  10.44 Gyr from the start, offset
1.0 in $x$ with each plot. The progenitor centers, marked by the circles, vary
because the sub-halos
induce small changes in its orbital period. }
\label{fig_xyz}
\end{figure}

\begin{figure}
\begin{center}
\includegraphics[angle=-90,scale=0.8]{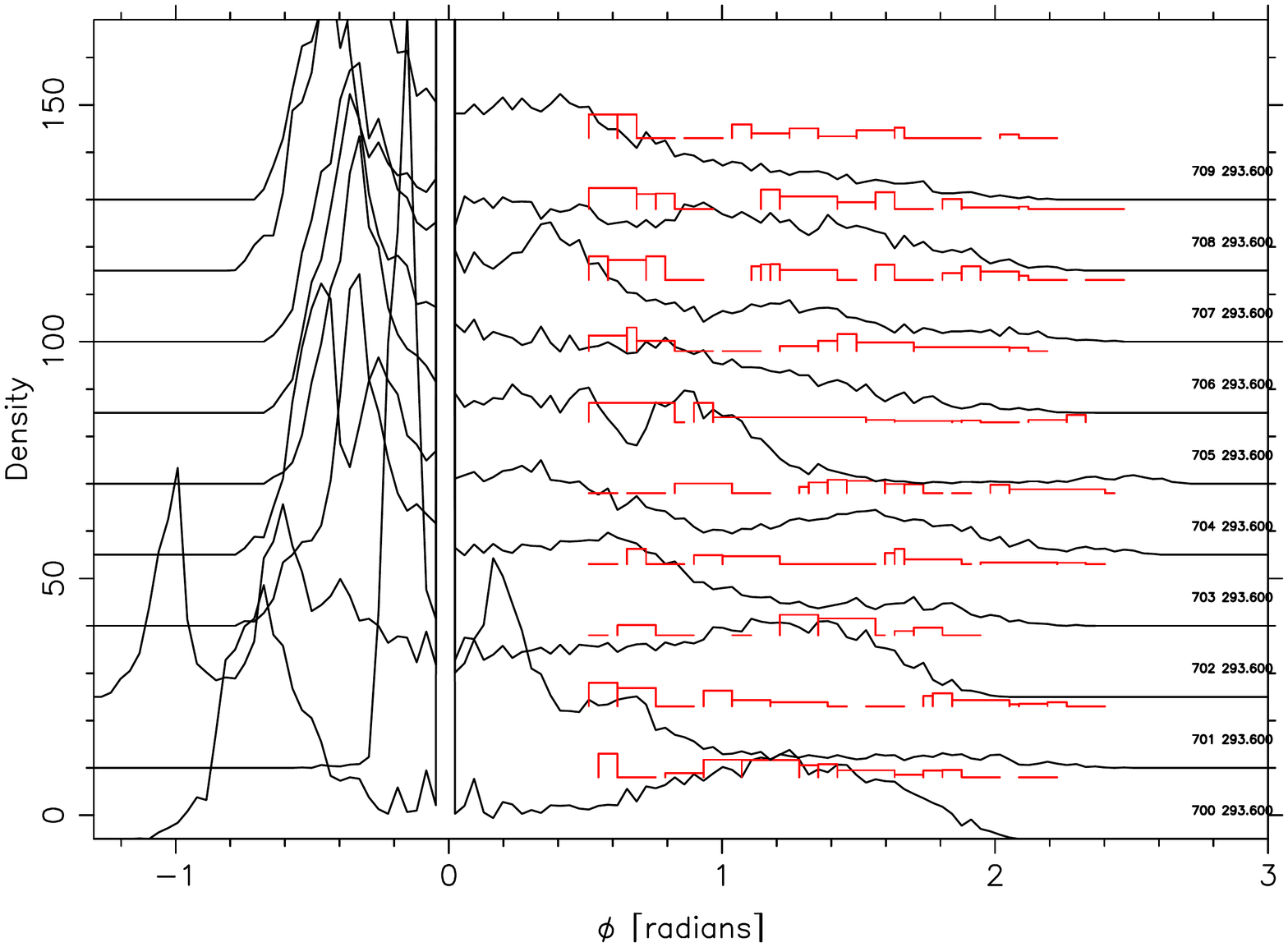}
\end{center}
\caption{The density along the stream for the ten realizations of the eccentricity 0.32 stream at 10.44 Gyr 
age, the same time as shown
in Figure~\ref{fig_xyz}. Angles are measured with respect to the progenitor which defines an angle of zero and has the large
density peak. 
The midpoint of the leading arm of the stream is near pericenter and near the solar
location of $x=1$, see Figure~\ref{fig_xyz}, roughly replicating 
the GD-1 observations. The stream is observed from a position offset from
the center, equivalent to the solar radius in the model. The gap analysis in this figure
is allowed to extend over larger angles than used for our GD-1 comparison. The large density peak at negative
angles (trailing arm of the streams)
 is the result of particle pileup at lower angular velocities at large radial distances, and, increased distance
from the viewing point. A very massive sub-halo has gone through the trailing arm of the stream second from the bottom.
The lowest density line shows the \citet{Kupper:12} epicyclic variations are only visible for about two cycles
away from the progenitor.
}
\label{fig_tenstreams}
\end{figure}

The GD-1 stream has an inferred orbital pericenter of approximately 14 and apocenter of 30 kpc. To fully populate our model with
sub-halos that have pericenters outside this range would be computationally inefficient. 
The radial distribution of sub-halos is smoothly diminished to zero at large radius 
with the function $\exp{[-{1\over 2}(r/r_g)^3]}$. Recalling
that the inferred apocenter for GD-1 is a little under 30 kpc,
$r_g$ is set at 7 scale radii, or 56 kpc, which leaves the inner region largely intact except for sub-halos
on very low probability very radial orbits,
but fairly quickly reduces the numbers at larger radii. 
The outcome is that within  245 kpc the sub-halo numbers are reduced 91\% with only a 4\% reduction inside 32 kpc. 
Sub-halos are randomly drawn from the resulting radial distribution and placed in a random spherical distribution. 
The equilibrium velocity dispersion of this distribution is calculated from Jeans' equation using the spherical halo
alone.  We then have radial density and velocity distribution function 
which we use to set up the population of sub-halos in an approximate equilibrium.
Although the calculation ignores the bulge and disk components of the inner potential
 the distribution is sufficiently close to equilibrium for our
purposes in the region of the GD-1 orbit, see Figure~\ref{fig_subhalos}, 
where the initial and final sub-halo distributions are visibly similar.

The mass distribution of sub-halos is
generated from the \citet{Aquarius} number-mass distribution, $N(>m)\propto m^{-0.9}$. The masses
are restricted to a maximum where they would have a short dynamical friction time, and, a minimum below which 
their effects on the stream are negligible. That is, 
the dimensionless mass range of sub-halos is $3\times 10^{-3}$ to $3\times 10^{-6}$, where one mass unit in MW2014 is
$9.006\times 10^{10} \msun$, or  $2.7\times 10^8$ to $2.7\times 10^5 \msun$.  
The total sub-halo mass contained to the minimum sub-halo mass allowed 
is 50\% of the total in the mass range down 
zero mass. These numbers vary from halo to halo and 
the numbers in the inner halo will vary depending on recent mergers. For the adopted normalization
there is typically one sub-halo generated above $m=10^{-3}$ ($9\times 10^7 \msun$), with a median
total number of about 120 with individual runs in a suite of 100 ranging from about 14 to 219. 

The star cluster is usually simulated with 50,000 particles, but some higher resolution results with 500,000 particles are
reported below. The system is evolved with the parallel shell code described in \citet{Carlberg:15a} which provides high dynamical 
accuracy at reasonably low cost. 
This code is, by design, collisionless and does not capture any of the complex collisional processes that can drive evaporation
in dense, low velocity dispersion, globular clusters. 
There are two reasons that a collisionless cluster is an acceptable approximation for our study. 
First, for clusters on eccentric orbits the rate of tidal heating occurs on an orbital time scale and 
generally exceeds the two-body heating of core collapse, an example being the relatively diffuse Pal~5 cluster and its associated stream.
Second, GD-1 has no known progenitor to use to determine the importance of two-body relaxation processes.

The MW2014 potential is the external potential in which the sub-halos orbit as test particles. The star cluster
is self-gravitating and responds to the sub-halos. The code accurately reproduces the results of a full n-body code for
situations of interest. 
All particles
are advanced with time steps of 0.002 units using a leap-frog integrator. The small time steps capture the internal dynamics of the
star cluster and the interactions between the stream and sub-halos.
The simulations run to time 300, 
where one time unit in MW2014 is $3.556 \times 10^7$ years, so a duration of 10.67 Gyr.

The star cluster is started on an orbit similar to that inferred for GD-1, with no attempt to replicate
it precisely.  
The starting point
is $x=3.6$ (28.8 kpc), $y=0$, $z=2.5$ (20 kpc), which is approximately the location of the
current orbital apocenter of the stream\citep{Willett:09, KRH:10}. 
The progenitor is started with a purely tangential velocity, $v_x=v_z=0$, $v_y$ of 0.85 of the local circular velocity calculated from
the MW2014 potential. 
The resulting orbit has an eccentricity of 0.32, essentially equal to  the 0.33 inferred for GD-1 \citep{Willett:09}. 
We also ran orbits with a starting velocity of 0.7 of circular, which leads to an orbital eccentricity of 0.44. 
A more eccentric orbit has the benefit of increased mass loss, roughly 55\% as opposed to the 24\% for the lower 
eccentricity, which provides better statistics in the stream. Although more eccentric orbits will generally blur out gaps 
more quickly we find that the gap spectrum is very similar for these two fairly similar orbits.

\subsection{Stream Realization-to-Realization Variations}

The stellar streams that develop  in the simulations need to be put into coordinates similar to those for GD-1
to make the gap distributions and the density alone the stream usefully comparable.
The density of stars in the stream varies along its length and with time if the progenitor is in an elliptical orbit \citep{Johnston:98}. 
Viewed from the center of the galaxy the mean density in angular coordinates peaks where the stream is moving slowly and 
the stream is far from the center, that is, at apocenter, and is low at pericenter. 
Having a long angular segment at fairly constant density requires a point of observation 
close to one of these locations, which for GD-1 is pericenter. We therefore 
analyze times when some portion of the stream is near pericenter and place a point of observation nearby. 
The nearly constant density profile of GD-1 is largely a remarkable coincidence of 
timing and location relative to the solar location, and of course it is much easier to find streams nearby.

The  position
and angular momentum of the progenitor cluster defines an instantaneous orbital
plane and a great circle around the sky.  The point of observation is at  the solar position, 
1 unit (=8 kpc) from the model center at a location
of [x,y,z] = [0.707,0.707,0], see Figure~\ref{fig_xyz}.
The stream is observed at time 293.6, or 10.44 Gyr, 
the time for the streams shown in Figure~\ref{fig_xyz}.  The stream
particle positions are projected onto the coordinate system for each
simulation and assigned angles relative to the progenitor center. 
At this time and position the leading arm of the stream appears to have an angular distance on the sky of nearly 180\degr.  
If observed from
the center the stream would be shorter and the density distribution would have a larger mean density variation along the stream. All of the results and plots here are measured in the coordinates of the solar position analog.

The densities  along the stream for ten realizations of the standard initial
conditions are shown in Figure~\ref{fig_tenstreams}, observed from the solar position analog. The primary difference
between the simulations is the distribution of sub-halos. In particular the heaviest dozen or so sub-halos have the
largest effects and can have very different orbits that may or may not lead to any intersections with the stream.
The variation from realization to realization is substantial. 
Consequently the prediction of the number and width of gaps along the
stream will be a statistical prediction.

\subsection{Sub-halo Mass Subsets and Stream Variations}

\begin{figure}
\begin{center}
\includegraphics[angle=-90,scale=0.8]{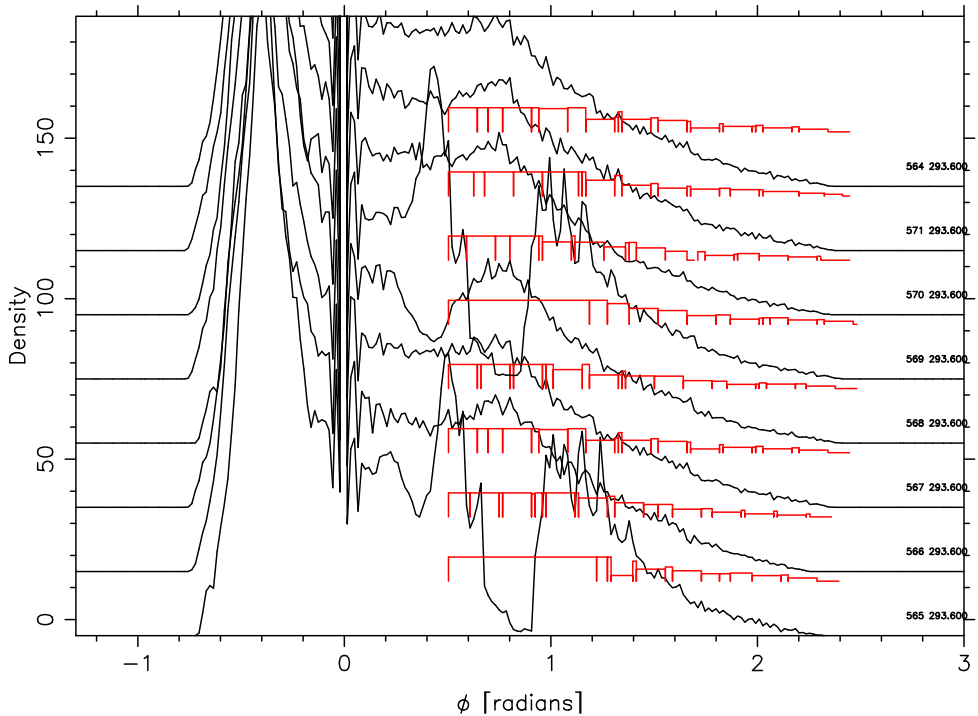}
\end{center}
\caption{A stream with a 500,000 particle progenitor displayed with all sub-halos (bottom)
and then rerun with sub-halos in
successively smaller mass ranges (see text), 
with the top density distribution being for a reference model with no sub-halos.
The mass ranges span a factor of 1000 in mass with approximately equal mass in each range.
In this case the large gap is the result of the 7 mid-mass range sub-halos, not a single large sub-halo.
}
\label{fig_satest}
\end{figure}

The effect of different mass ranges of sub-halos is displayed in Figure~\ref{fig_satest} for one 
of the standard stream simulations.
The line at the bottom of the 
plot is the density along the leading
arm of the stream for the 101 sub-halos over the whole mass range from $3\times 10^{-3}$ to $3\times 10^{-6}$.
($2.7\times 10^8$ to $2.7\times 10^5 \msun$).
Second from the bottom is the mass range $3\times 10^{-3}$ to $1 \times 10^{-3}$ 
($2.7\times 10^8$ to $9\times 10^7 \msun$), in which there is one sub-halo. 
Typically there is a single halo is present in this mass range 
and it hits the stream about half the time. In the mass range $1\times 10^{-3}$ to $3 \times 10^{-4}$ 
($9\times 10^7$ to $2.7\times 10^7 \msun$)
there are no  sub-halos, 
$3\times 10^{-4}$ to $1 \times 10^{-4}$ 
($2.7\times 10^7$ to $9\times 10^6 \msun$)
has 4, $1\times 10^{-4}$ to $3 \times 10^{-5}$
($9\times 10^6$ to $2.7\times 10^6 \msun$)
 has 7, $3\times 10^{-5}$ to $1 \times 10^{-5}$
($2.7\times 10^6$ to $9\times 10^5 \msun$)
has 27, and at the top $1\times 10^{-5}$ to $3 \times 10^{-6}$
($9\times 10^5$ to $2.7\times 10^5 \msun$)
 has 62. 
These numbers are for the restricted inner halo sub-halo population, but do reflect that for the MW2014 potential's dark
matter sub-halo the numbers of dynamically relevant sub-halos are not large.
The simulations show how the overlap of sub-halo impacts on the stellar stream makes individual sub-halo
hits very difficult to discern unless the sub-halo is very massive. Even then, smaller mass sub-halos, those 
near $m=10^{-4}$, or $10^7 \msun$, can create significant
structure on the density profile as is dramatically shown in Figure~\ref{fig_satest}, where the 7 halos that are individually about 3\% of the maximum allowed mass, create a gap that could be mistaken for the passage of a single much more massive halo. 

Smaller mass sub-halos, those below $1\times 10^{-5}$, or $10^6\msun$, second and third
from the top in Figure~\ref{fig_satest}, produce projected density variations so small 
that they are barely distinguishable from the no sub-halo case, the top line in  Figure~\ref{fig_satest}.
Figure~\ref{fig_satest} shows that the combined effect of sub-halos is not a simple sum in density space of the
outcomes of individual sub-halos. That is the density (and velocity)
fluctuations in a stream as a result of population of sub-halos is substantially non-linear, even though the changes in
orbital momentum variables (actions) of individual particles remain effectively in the linear regime.
As a consequence
it will be very difficult, if not impossible, to infer the properties of individual lower mass sub-halos 
even with full phase space information for the stars in the stream.
Massive sub-halos, 
those up around $10^8 \msun$ ($10^{-3}$ in our dimensionless units) are relatively rare and cause
large perturbations in a stream so offer an opportunity for individual study although even their 
gaps are perturbed by the lower mass sub-halos that are present.

\section{Gaps in GD-1 like simulations}

\begin{figure}
\begin{center}
\includegraphics[angle=-90,scale=0.6,trim=93 0 90 0, clip]{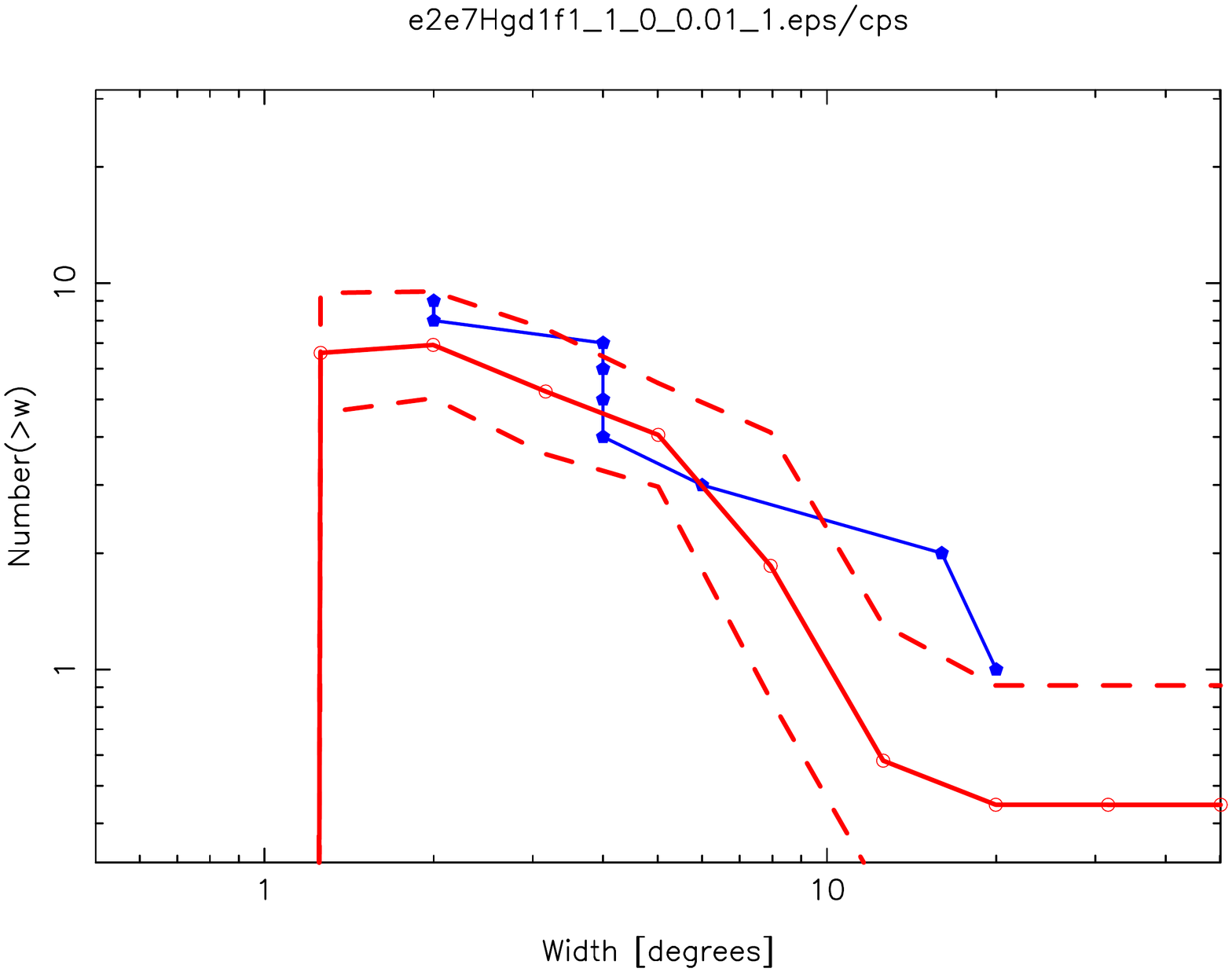}
\put(-442.5,-237){\includegraphics[angle=-90,scale=0.726,trim=93 0 0 0, clip]{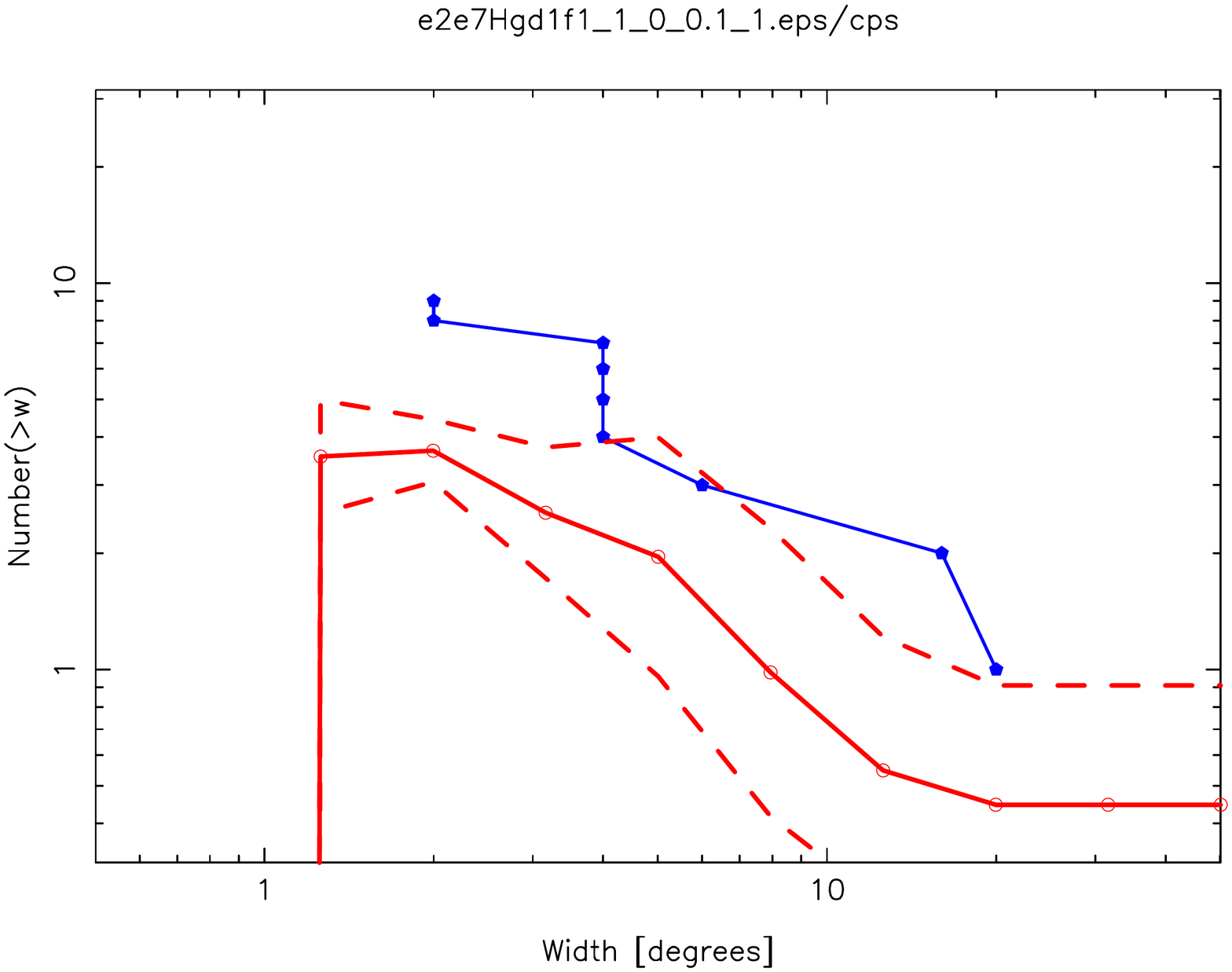}}
\put(-222,-240){\includegraphics[angle=-90,scale=0.31,trim=93 0 0 0, clip]{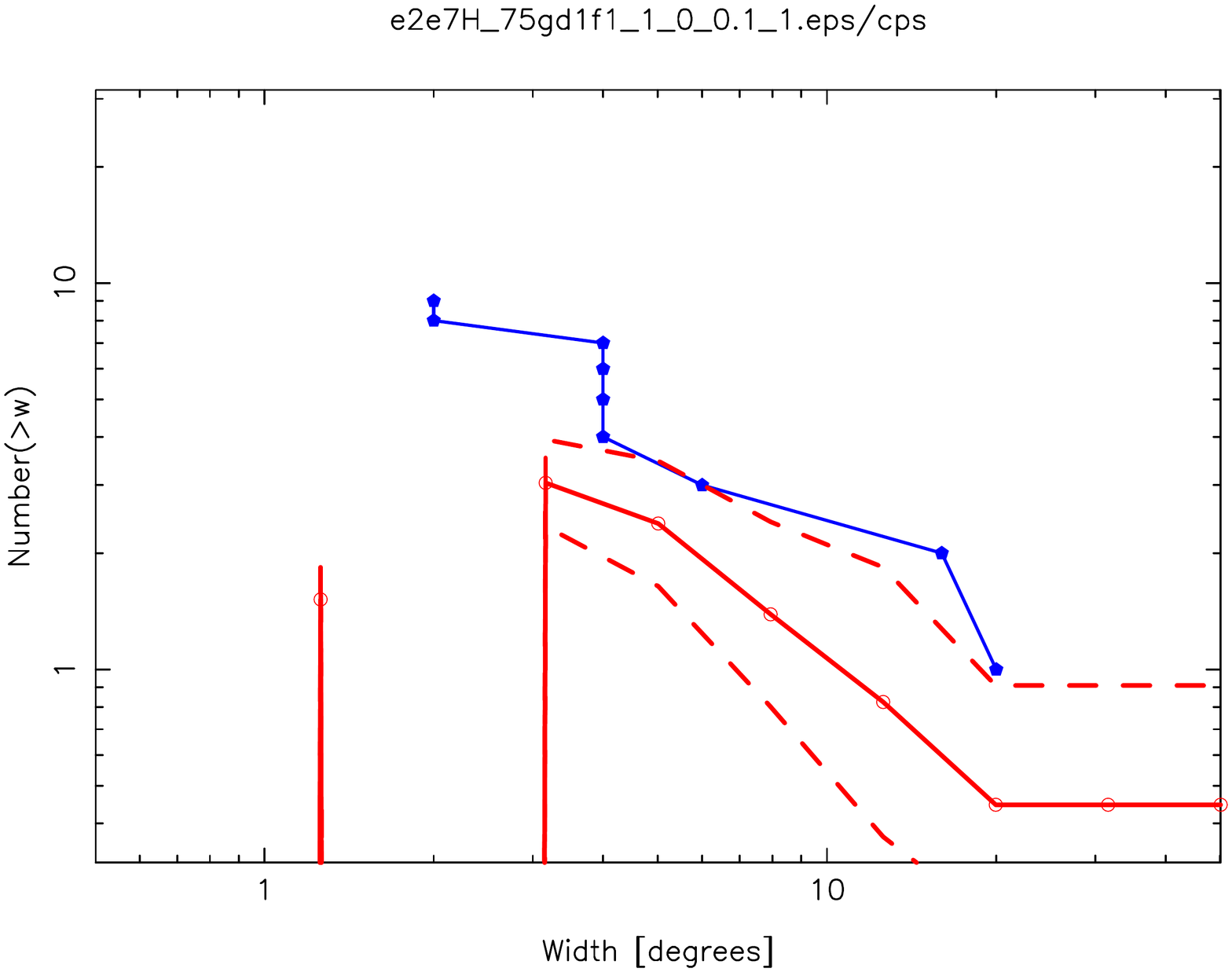}}
\end{center}
\caption{The gap distribution found with the $w_1$ filter 
averaged over 100 simulated streams at time 293.6, 10.44 Gyr, at full resolution with the 
RMS deviation from the mean plotted above and below the mean as dashed lines. The 
same filter applied to GD-1 is shown as (blue) line with points.
The minimum gap height is 0.01 in the top panel and 0.1 in the bottom panel.
The inset figure in the lower panel shows the same analysis at time 154,  5.48 Gyr, requiring a minimum gap height of 0.1.
 }
\label{fig_ng1}
\end{figure}

One hundred realizations of  clusters on the  GD-1 like orbit in the MW2014 potential are run for 300 time
units, or 10.67 Gyr . We focus on time 293.6, 10.44 Gyr, which is the last 
orbit when the leading arm of the stream is in a
configuration somewhat similar to GD-1, where the stream is near
pericenter and near the sun.

\citet{Kupper:08,Kupper:12} demonstrate that as the gravitational tide of the host galaxy pulls stars
away from a globular cluster that coherent epicyclic orbital motions of the stars leads to density variations in the
tidal stream.  These density variations naturally appear in the simulations here as well but are only
important near the progenitor system as is visible in Figure~\ref{fig_tenstreams}.
We do not want to confuse these variations with sub-halo density variations, so the measurements of our simulations 
exclude the region near the progenitor, specifically the 40\degr\ of the stream closest to the progenitor. Recall that
GD-1 has no visible progenitor, so this exclusion region is appropriate. 

The offset viewing point near pericenter
leads to densities along the stream are relatively constant with azimuthal angle, similar to GD-1.  
The particles are placed into density bins the same size as used 
in the GD-1 analysis, usually 2\degr. The resulting density along the stream is passed through a Gaussian 
high pass filter that removes the large scale density variations of the stream so that the emphasis is on
the gaps, not the large scale density variations.

The number of gaps depends on the minimum height threshold used.
A minimum height of 0.1 gives all the gaps of the real GD-1. A ``perfect'' noiseless gap that exactly matches
the filter shape and has a depth such that the density goes to zero at the bottom would give a
gap of height  of  exactly 1.0. 
We use gap height thresholds of 0.1, which selects the more robust features of the simulations and includes all of the
GD-1 gaps that the algorithm finds, and a low threshold of 0.01,
 which usually finds all the  gaps to be found in the simulations.

\subsection{Gaps in Low Noise Streams}

First we examine simulated streams with about a factor of 3 to 4 higher signal to noise in the
currently available GD-1 data. 
In Figure~\ref{fig_ng1} in the top panel the minimum gap height allowed is 0.01 which collects all the gaps in 
the simulations at time 293.6, 10.44 Gyr. 
The displayed quantities are the mean and the 1$\sigma$ confidence interval which are calculated
from the standard deviation of the counts.
Also shown in  Figure~\ref{fig_ng1}
is the GD-1 gap distribution measured with precisely the same parameters.
The numbers of small separation gaps increase significantly 
as the minimum gap height counted decreases, indicating that
noise is a major factor for their numbers. The numbers of large gaps in the simulations are not very sensitive to
the minimum gap height. Moreover, there is generally good agreement with the GD-1 measurements of larger
gaps. 

The number of gaps depends on the minimum gap height allowed for counting. 
The lower panel of Figure~\ref{fig_ng1} shows the gap distribution with a minimum height counted of 0.01, showing
that most of the small gaps are relatively small amplitude gaps but that the bigger gaps are relatively high amplitude.
A set of these gap size distributions for a subset of ten simulations is
shown in from Figure~\ref{fig_tenstreams}.  
The  inset in the lower panel of Figure~\ref{fig_ng1} 
shows the mean gap distribution at time 154, 5.48 Gyr, about half the age of the other plots, for the 0.1 minimum gap height.
The large gaps are, on the average, already at the mean level 
seen at the end of the simulations, although the spread of values about the mean is somewhat smaller.
That is, the gap distribution becomes stable at about half the Hubble age, 
when sufficient stream length is available and as new gaps are created older
ones fade away.  This is expected since the lifetime of smaller gaps in a stream with a GD-1 like
orbit is typically 5 to 6 orbits, or about 3 Gyr, \citep{Carlberg:15a}.

\begin{figure}
\begin{center}
\includegraphics[angle=-90,scale=0.8]{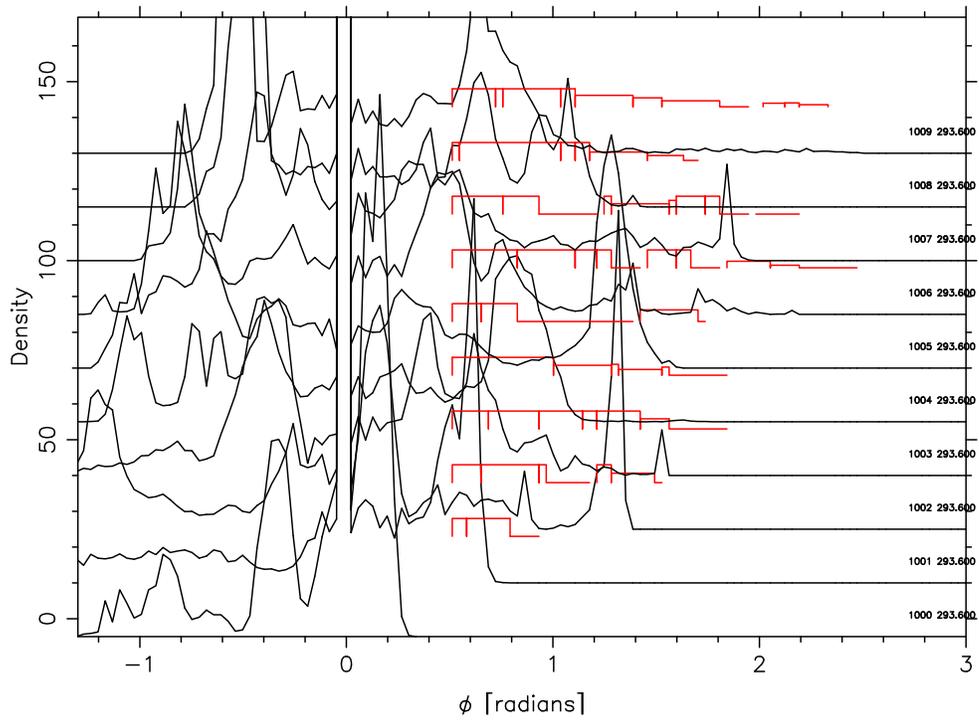}
\end{center}
\caption{Same as Figure~\ref{fig_tenstreams} except for a sub-halo mass fraction 
raised a factor of ten, leading to ten times the number of sub-halos. The
vertical scale has been kept the same as the earlier figure to emphasize the 
how extreme the density variations are. The angular phase differences between simulations are also much larger.
}
\label{fig_tenten}
\end{figure}

\begin{figure}
\begin{center}
\includegraphics[angle=-90,scale=0.7,trim=93 0 0 0, clip]{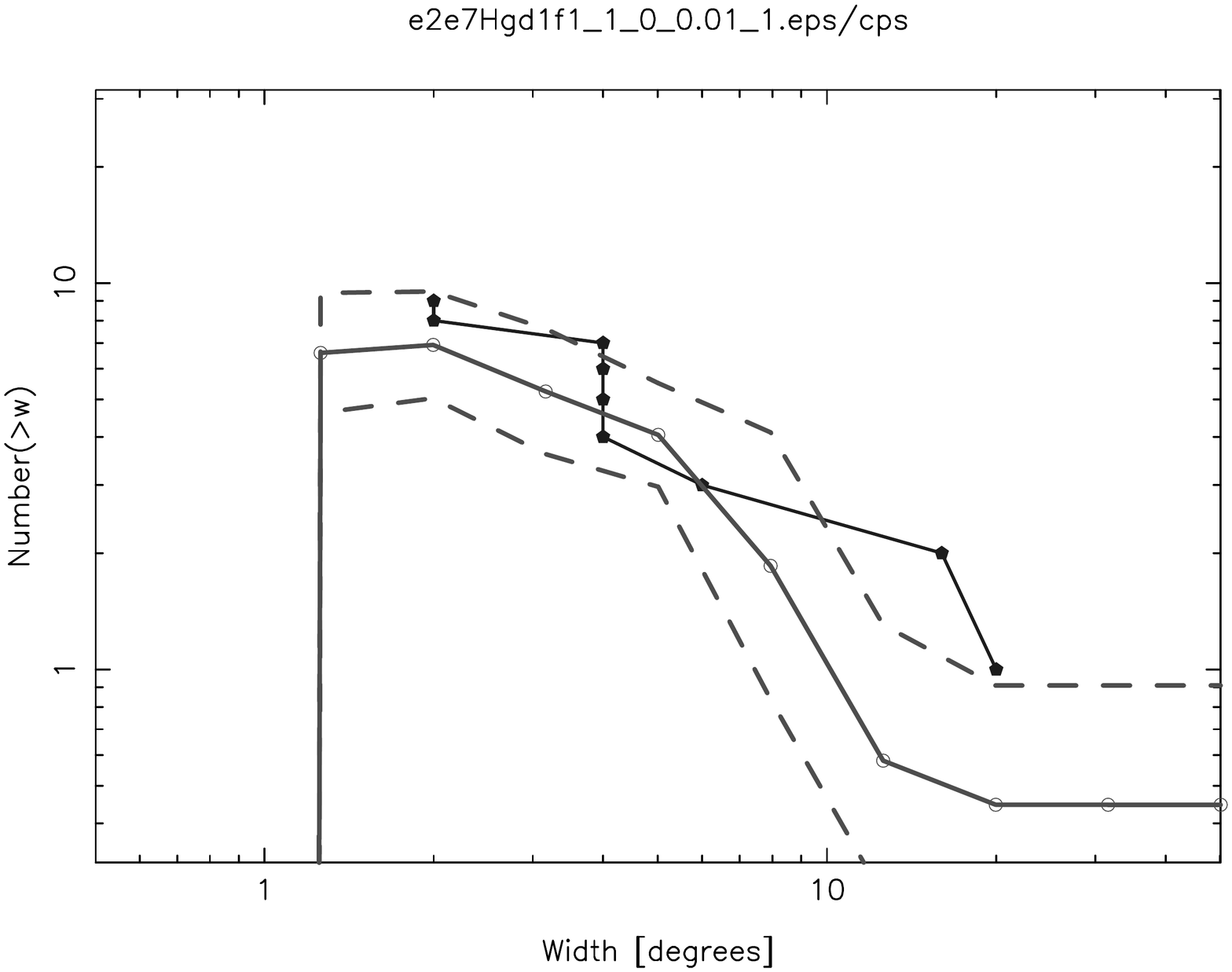}
\put(-426.5,0){\includegraphics[angle=-90,scale=0.7,trim=93 0 0 0, clip]{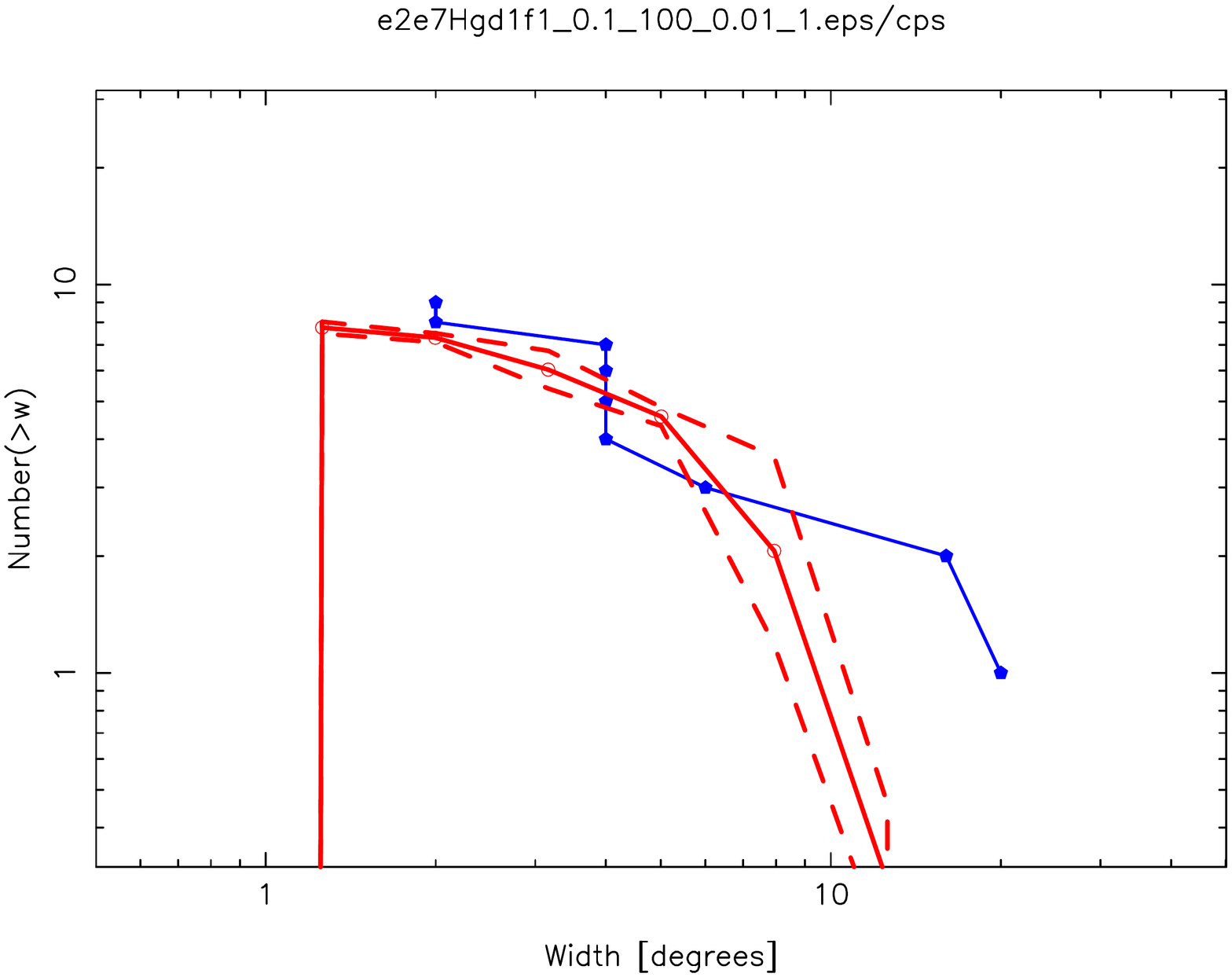}}
\end{center}
\caption{Same as Figure~\ref{fig_ng1} but for streams that are pure noise for a 0.1 gap height
threshold. Comparison
to Figure~\ref{fig_ng1} (top panel, shown in grey here) shows that a real stream has significantly more gaps 
at 10\degr\ and more.  }
\label{fig_ng0}
\end{figure}

We also explore the outcome with the mass fraction in sub-halos increased.
Doubling the mass fraction in sub-halos doubles the number of large sub-halos intersecting the streams,
but they remain sufficiently infrequent that the overall distribution remains about the same.
For an extreme effect, the mass in sub-halos
is increased a full factor of 10 which boosts the active sub-halo mass fraction in the inner region to 1.9\% from 0.19\%
 so the simulation
remains physically sensible, although perhaps cosmologically implausible except during major accretion events.
In this case there are so many large sub-halo hits that the stream always displays large deep gaps with a range of density
contrast about 3 times larger than those seen in the simulations with the standard parameters. These
large density variations do not appear to exist in the GD-1 data.
The extreme density structures shown in Figure~\ref{fig_tenten} are 
not very successfully recovered with the current gap filters.

\subsection{Gaps in Noisy Streams}

Figure~\ref{fig_ng0} shows the gap spectrum for essentially pure noise, created by adding a Gaussian random number 
to the stream densities sampled at the rate of one particle out of one hundred. 
In addition to the mean and 1$\sigma$ spread, the circles show 
the result of the individual measurements of the 100 streams. 
The noise explains all the gaps smaller than about 8\degr, but
falls short of the gaps greater than 10\degr.  
It is also notable that the pure noise gap distribution has a much smaller
spread of distributions around the mean compared to the stream results. 
Quantitatively, gaps of 10\degr\ and larger are less than 1\% probable from noise. That is,
at 99\% confidence the large gaps are not due to noise alone.
We now can claim to understand  the nature of the excess variance in the GD-1 stream: it is due to
an excess of density variations that match the  shape of gaps over scales of 10\degr\ and larger. 

\subsection{GD-1 Noise matched Streams}

\begin{figure}
\begin{center}

{\includegraphics[angle=-90,scale=0.7,trim=93 0 0 0, clip]{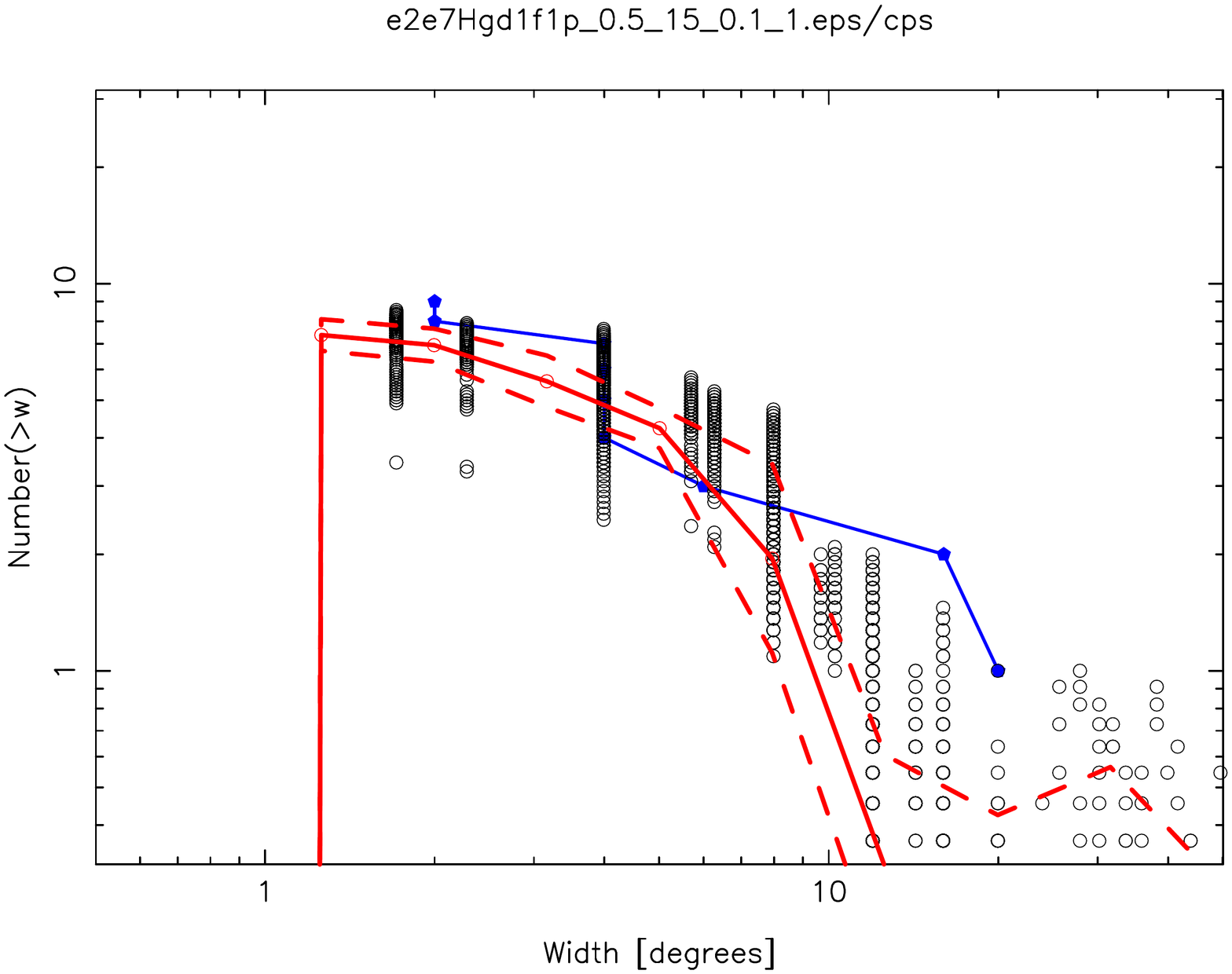}}
\put(-190,-5){\includegraphics[angle=-90,scale=0.25,trim=93 0 0 0, clip]{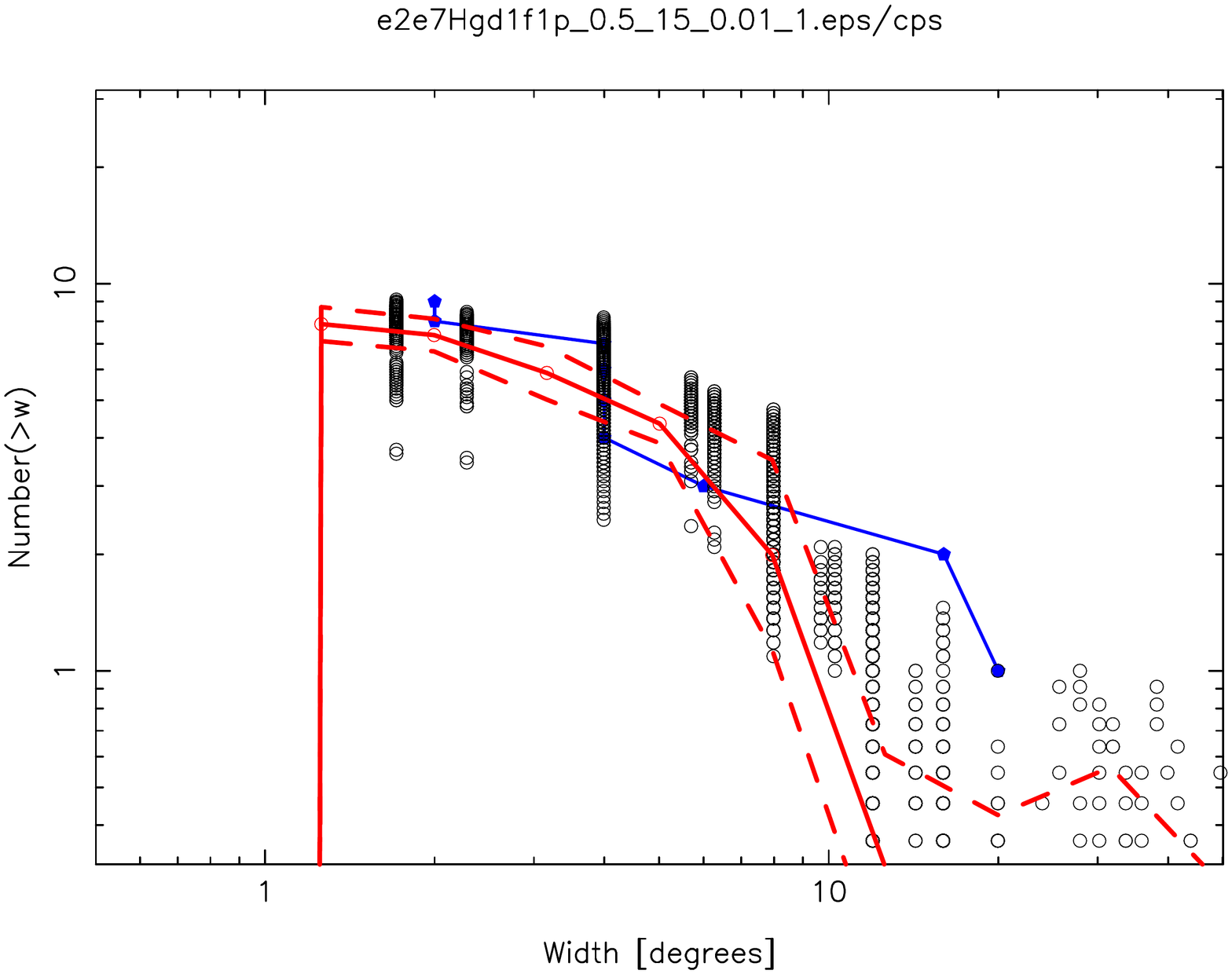}}
\end{center}
\caption{Same as Figure~\ref{fig_ng1} but for streams that are randomly sampled to 50\% of their density, with
noise added to  approximate the signal to noise of 
GD-1. The gap height threshold is 0.1. The inset shows the results with a 0.01 gap height threshold which is essentially 
identical to the higher threshold version. The circles show the results for each simulation, so the large angle gaps
are in excess of the standard sub-halo predictions.
}
\label{fig_ng2}
\end{figure}

Neither Figure~\ref{fig_ng1} nor Figure~\ref{fig_ng0} have the statistics of the GD-1 stream, where 
the estimated s/n of a density bin is 2.3. The simulations have a mean s/n per 2\degr\ bin of 8.5, having higher
numbers of particles than stars in the current data and no foreground noise. 
To create  a stream with statistical properties comparable to GD-1 we 
first randomly select a subset of the particles from the each simulation used
in Figure~\ref{fig_ng1} and then add Gaussian noise to
reflect the noise in the foreground/background distribution.
There are 
about 10,000 particles lost from each simulations. Half of those particles are in the leading arm of the stream.
\citet{GD:06} estimated the number of stars in their 63\degr\ stream segment as $1800\pm 200$. We are
using the 73\degr\  segment from \citet{CG:13} which is longer but probes the
same depth in the sky, so  estimate that there are about 2500 stars. Therefore the stream data
has about half the numbers of stars as particles in  the simulation streams. 
To allow for the noise in the foreground which has been subtracted we 
add a Gaussian noise that corresponds to a background of 225 units to each density bin.
The outcome is that the sub-sampled, noise-added simulation data have a mean signal-to-noise of 2.3, equal to that for GD-1. 
We then measure the gaps, 
repeating the sampling and noise addition process 11 times on
an individual stream to create an average. 
The result is displayed in Figure~\ref{fig_ng2}. 
The outcome is close but definitely not identical to the result for a pure noise spectrum
since there are more large, 10\degr\ and up, gaps than
pure noise. However the noise matched stream gap distribution falls somewhat short of what is seen in GD-1
for large gaps.
This may be a statistical fluctuation, or, it could point to 
some systematic error in the data at larger scales. Or, there could be somewhat more large sub-halos that 
expected in a virialized LCDM galaxy halo.
Although the excess of large gaps relative to the LCDM prediction is
interesting and could call for an enhanced sub-halo population in the inner halo, we will examine this 
issue again as better data become available.

\section{Discussion and Conclusions}

The gap filtering approach is a decomposition of the density along a stellar stream, $d(\phi)$, 
to a filtered density field $D(\phi,g)$, where $g$ is related  to the length of the gap, here
${\rm w}\le 6g$. The filtered distribution
is then examined to find the local peaks, after which we  find the $g$ value which provides the highest
peak at a given $\phi$.  
The outcome is a set of gaps along the stream that are characterized with their width and the minimum height required of the gap.
The cumulative distribution of gap sizes, $N(>{\rm w})$, is a way of characterizing the density fluctuations 
in the stream.

To develop insight into this distribution we do identical measurements on a suite of  
GD-1 like streams that develop  in a Milky-Way potential with and without sub-halos.
The distribution of gaps sizes stabilizes after about a half dozen orbits, 3 Gyr, so the density variations 
are not very sensitive to the age of the stream or precisely how much mass is in the sub-halos. 
In comparison to a stream that is
simply a random density field, the GD-1 stream has a significant excess of large gaps. 
That is, what the filters find to
be  large gaps are the source of the excess variance in the GD-1 stream above the noise level. 
However, if the simulations have their signal to noise artificially reduced to the level of the GD-1 stream,
the simulations do not produce enough large gaps to match the data, although the discrepancy is not large.
One possible explanation is that we have over-estimated the random errors which are added to the simulation data,
or, there are unresolved 10 degree scale systematic errors in the measurement of the GD-1 density distribution. 
Another possibility is that the sub-halo mass distribution has relatively more massive sub-halos in the
standard Aquarius halo which we use to normalize our results.

The filters used here are simple first steps, adequate to indicate the nature of the excess variance in GD-1 given
the current data.  Future data with improved signal to noise for density measurements
along the stream will merit a more thorough exploration of filters.  
In particular, the current data has signal-to-noise of 2.3 in the 2\degr\ density bins, which at the same depth will rise to about 6.6 
as foreground stars are removed with new astrometric data and ever-improving photometric techniques.  
Slightly increased depth will then put the  signal-to-noise in the regime of our noise-free simulations and increase 
the confidence of gap finding to their level.
A wider range of shapes will allow much better matches
to the shapes of the gaps and provide more confident detections. 
The signal-to-noise of observational data will improve a lot
as proper motions help remove foreground stars and deeper images add more stream stars. 
A future step is to add stream star velocities to the analysis. 

\acknowledgements

This research was supported by CIFAR and NSERC Canada.

\end{document}